\documentclass[a4paper,english,10pt,superscriptaddress,reprint,amsmath,amssymb,aps,twocolumn,floatfix,notitlepage]{revtex4-2}
\allowdisplaybreaks[2]
\usepackage{graphicx}
\usepackage{bm}
\usepackage{dcolumn}
\usepackage[colorlinks=true,citecolor=darkred,urlcolor=darkred, pdfborder={0 0 0}]{hyperref}
\usepackage{setspace}
\usepackage{caption}
\usepackage{subcaption} 
\usepackage{slashed}
\usepackage[normalem]{ulem}
\usepackage{float}
\usepackage[usenames,dvipsnames]{xcolor}  
\usepackage{microtype}
\usepackage{lipsum}
\usepackage{amsmath}
\usepackage{amssymb}
\usepackage{mathrsfs}
\usepackage{placeins}
\definecolor{darkred}{rgb}{0.6,0,0}
\definecolor{linkcolor}{rgb}{0,0,0.5}
\usepackage[T1]{fontenc} 
\usepackage{array}
\usepackage{pifont} 

\definecolor{linkcolor}{rgb}{0,0,0.5}
\usepackage{orcidlink}
\begin{document}
\begin{flushright}
OU-HET-1251
\end{flushright}
\title{\boldmath \color{BrickRed} Triple $Z'$ signatures at $Z$ factories}
\author{Takaaki Nomura \orcidlink{0000-0002-0864-8333}}
\email{nomura@scu.edu.cn}
\affiliation{College of Physics, Sichuan University, Chengdu 610065, China}
\author{Kei Yagyu \orcidlink{0000-0001-6382-6043}}
\email{yagyu@het.phys.sci.osaka-u.ac.jp}
\affiliation{Department of Physics, Osaka University, Toyonaka, Osaka 560-0043, Japan}
%
%
\begin{abstract}
\noindent
We discuss triple $Z'$ boson signatures via the decay chain of $Z \to Z' \phi \to Z' Z' Z'$, with a new light scalar $\phi$, at future Z factories such as CEPC and FCC-ee. These new bosons $\phi$ and $Z'$ naturally appear in models with a new $U(1)$ gauge symmetry which is spontaneously broken and introduced in various new physics scenarios. 
The branching ratio of $Z \to Z' \phi \to Z' Z' Z'$ can be larger than $10^{-12}$, which gives $O(1)$ events at Tera-Z experiments, when a product of $g_X^{}$ (new gauge coupling) and $\zeta$ ($Z$-$Z'$ mixing) is larger than around $10^{-6}$. 
We find that the search for $Z \to Z'Z'Z'$ can significantly improve the current bound on a kinetic mixing parameter $\epsilon$ in the dark photon case, where $e \epsilon \gtrsim \mathcal{O}(10^{-5})$ with $g_X^{}={\cal O}(1)$ can be explored at Tera-Z experiments. 
We also show that a sufficiently large number of events with multi-lepton plus hadronic jets can be obtained in benchmark points, which cannot be realized by the usual decay of Z in the standard model. 
\end{abstract}
\maketitle 

\section{Introduction}

The gauge principle has successfully described fundamental forces, e.g., strong force, weak force, electromagnetic force and probably gravity as well~\cite{Utiyama:1956sy}. A natural question that follows is, 'Is there a fifth force in nature?'. 

A new $U(1)$ gauge symmetry, which we denote as $U(1)_X$, is quite often introduced in various physics motivations, which deduces the fifth force in a simple way. For instance, grand unified theories, such as $E_6$ and $SO(10)$ models, predict $U(1)$ gauge symmetries in a low-energy effective theory~\cite{Hewett:1988xc}.     
In addition, the $U(1)_X$ symmetry naturally introduces right-handed neutrinos because of the anomaly-free condition~\cite{Appelquist:2002mw}. Furthermore, a spontaneous breakdown of the $U(1)_X$ symmetry can be origin of tiny neutrino masses, stability of dark matter and baryon asymmetry of the Universe, see e.g.,~\cite{Matsui:2023bwa}. 
Therefore, testing the fifth force is quite important to narrow down new physics scenarios. 

As a common feature, a $Z'$ boson, i.e., a new neutral gauge boson, appears in models with $U(1)_X$, so that the detection of $Z'$ turns out to be direct test of such a scenario. 
Searches for $Z'$ have been performed in various experiments, and so far no clear evidence of the discovery of $Z'$ has been reported, by which 
the mass $m_{Z'}$ and the new gauge coupling $g_{X}^{}$ (or the kinetic mixing parameter $\epsilon$ especially for the dark photon case) have been constrained. 
In particular, the range of $m_{Z'}\lesssim 100$ MeV has been highly constrained to be $|\epsilon| \lesssim 10^{-9}$ by the combination of the beam dump experiments~\cite{Riordan:1987aw,Bjorken:1988as,Batell:2014mga,Marsicano:2018krp,Blumlein:2011mv,Blumlein:2013cua,Gninenko:2012eq} and observation of the supernova SN1987A~\cite{Chang:2016ntp} in the dark photon model, except for the region $10^{-5} \lesssim |\epsilon| \lesssim 10^{-3}$ and $10~\text{MeV} \lesssim m_{Z'} \lesssim 100$ MeV. 
On the other hand, the region with $m_{Z'} \gtrsim$ 100 MeV is less constrained to be 
$|\epsilon| \lesssim $ $10^{-3}$-$10^{-4}$. 
See Ref.~\cite{Fabbrichesi:2020wbt} for the summary of the current constraints in the dark photon model. 
Similar constraints on $g_X^{}$ can also be applied to the other $U(1)_X$ models such as the $B-L$ model, but additional constraints, e.g., the Texono experiment~\cite{TEXONO:2009knm}, can appear because of the non-vanishing interaction between $Z'$ and neutrinos. 

Another common feature is the appearance of a new Higgs boson $\phi$ which is associated with the spontaneous breaking of the $U(1)_X$ symmetry. 
Because both the masses of $Z'$ and $\phi$ are originated from the Vacuum Expectation Value (VEV) of $\phi$, these masses are naturally given to be in a similar region. 
Thus, for $m_{Z'}={\cal O}(10)$ GeV or smaller, 
$Z'$ can also be searched via the decays of the discovered Higgs boson, i.e., $h \to Z'Z'$ and/or $h \to \phi\phi \to 4Z'$ followed by the new Higgs boson decay $\phi \to Z'Z'$~\cite{Gopalakrishna:2008dv,Davoudiasl:2013aya,Chang:2013lfa,Curtin:2013fra,Falkowski:2014ffa,Curtin:2014cca,Izaguirre:2018atq,Nomura:2024pwr}. 
From the searches for multi-lepton events at LHC, we obtain the constraint on $g_X^{}$ and the mixing angle between $h$ and $\phi$, which can provide a stronger limit on $g_X^{}$ compared with the flavor constraints depending on the size of the mixing angle~\cite{Nomura:2024pwr}.  

In this Letter, we propose the possibility of exploring the region weakly constrained by the current experiments, i.e., 
$100~\text{MeV}\lesssim m_{Z'} \lesssim 10~ \text{GeV}$ at $Z$ factory experiments such as the Circular Electron-Positron Collider (CEPC)~\cite{CEPC-SPPCStudyGroup:2015csa} and the Future Circular Collider (FCC-ee)~\cite{FCC:2018byv}. 
Remarkably these future $Z$ factrory experiments are expected to produce $\mathcal{O}(10^{12})$ number of $Z$ boson (Tera-$Z$ experiments).
The $Z$ factory experiments are kind of an upgraded version of the LEP/SLC experiments. In particluar, $\sim 20$ million $Z$ bosons have been generated at LEP-I, and it has contributed to testing the electroweak interaction precisely~\cite{ALEPH:2005ab}.  
Now, we can consider the application of the $Z$ factories to explore new physics~\cite{CEPCStudyGroup:2018ghi,Liu:2017zdh}. We focus on a new decay channel, i.e., $Z \to Z'\phi \to Z'Z'Z'$, which can be realized even for a smaller scalar mixing. Thus, the triple $Z'$ production via the $Z$ decay can be complementary to the aforementioned multi-lepton searches at LHC. 
 
\section{Model \label{sec:models}}

We recapitulate a model with an extra $U(1)_X$ gauge symmetry in the minimal extension of field contents without introducing any exotic fermions~\footnote{Right-handed neutrinos are required to cancel the gauge anomaly associated with $U(1)_X$, and to generate active neutrino masses depending on the assignment of the $U(1)_X$ charges. In the following analysis, right-handed neutrinos are not relevant.}. 
The scalar sector is then composed of the isospin doublet Higgs field $H$ and an additional complex scalar field $\Phi$ which is singlet under the SM gauge symmetry, and carries the $U(1)_X$ charge $\tilde{X}_\Phi$. \footnote{The anomaly-free condition restricts the number of free $U(1)_X$ charges for all the fields in the $U(1)_X$ model to be 2 (3) if we (do not) introduce the Majorana mass term via the spontaneous breaking of $U(1)_X$  and flavor universal charge assignments. }

First, we consider the Higgs potential. The most general form is given by 
\begin{align}
V = & -\mu_H^2 |H|^2 -\mu_\Phi^2|\Phi|^2 + \frac{\lambda_H}{2}|H|^4 
\nonumber \\
& + \frac{\lambda_\Phi}{2}|\Phi|^4 + \lambda_{H\Phi}|H|^2|\Phi|^2. 
\end{align}
The mass squared parameters $\mu_H^2$ and $\mu_\Phi^2$ can be eliminated by solving the tadpole conditions for $H$ and $\Phi$ assuming nonzero VEVs of the Higgs fields  $\langle H\rangle = v/\sqrt{2}$ and $\langle \Phi\rangle = v_\Phi^{}/\sqrt{2}$. 
The CP-even components $\tilde{h}$ and $\tilde{\phi}$ from $H$ and $\Phi$, respectively, are mixed via the $\lambda_{H\Phi}$ term.
Their mass eigenstates are defined as
\begin{align}
\begin{pmatrix}
\tilde{h} \\
\tilde{\phi}
\end{pmatrix}=
R(\alpha)
\begin{pmatrix}
h \\
\phi
\end{pmatrix}, \ 
R(\theta) \equiv 
\begin{pmatrix}
\cos\theta & -\sin\theta \\
\sin\theta & \cos\theta
\end{pmatrix}
\end{align}
with the mixing angle $\alpha$ given by
\begin{align}
\tan2\alpha = \frac{2\lambda_{H\Phi}vv_\Phi}{\lambda_H v^2- \lambda_\Phi v_\Phi^2}. 
\end{align}
The scalar bosons $h$ and $\phi$ are identified as the discovered Higgs boson with the mass $m_h(\simeq 125~\text{GeV})$
and the additional one with the mass $m_\phi$, respectively.

Next, we discuss the gauge sector of our model. In general, we have a kinetic mixing between $U(1)_Y$ and $U(1)_X$ gauge fields written as
\begin{align}
\mathcal{L}  \supset & -\frac{1}{4} B_{\mu \nu} B^{\mu \nu} -\frac{1}{4} X_{\mu \nu} X^{\mu \nu}  - \frac{\sin \epsilon'}{2} B_{\mu \nu} X^{\mu \nu},  
 \label{eq:U1} 
\end{align}
where $X_{\mu\nu}$ and $B_{\mu\nu}$ are 
the field strength tensors for the $U(1)_X$ gauge field $X_\mu$ and the $U(1)_Y$ gauge field $B_\mu$, respectively.
The kinetic term can be diagonalized applying the following transformation~\cite{Holdom:1985ag};
\begin{equation}
\left(\begin{array}{c}
X_\mu\\
B_\mu\\
\end{array}\right)=\left(\begin{array}{cc}
{\rm sec} \, \epsilon' & 0 \\
-\tan \epsilon'  & 1 \\
\end{array}\right)\left(\begin{array}{c}
\tilde{Z}^\prime_\mu\\
\tilde{B}_\mu\\
\end{array}\right).
\label{eq:kinetic}
\end{equation}
The transformed field $\tilde{B}_\mu$ is further written by $\tilde{B}_\mu = c_W^{} A_\mu - s_W^{} \tilde Z_\mu$ ($c_W^{} = \cos\theta_W$, $s_W^{} = \sin\theta_W$), where $A_\mu$ is the photon field, $\tilde{Z}_\mu$ is the neutral gauge field and $\theta_W$ is the Weinberg angle. Two gauge fields $\tilde{Z}_\mu$ and $\tilde{Z}_\mu^\prime$ are mixed with each other after the spontaneous breakdown of the electroweak symmetry. 

Now, the covariant derivative for a generic field $\Psi$ is expressed as 
\begin{align}
D_\mu\Psi  \supset & \left[\partial_\mu  
-ig_Z^{}(T_\Psi^3 - s_W^2Q_\Psi)\tilde{Z}_\mu -ig_X^{}X_\Psi \tilde{Z}_\mu' \right]\Psi, \label{eq:cov}
\end{align}
where $g_X^{}$ is the $U(1)_X$ gauge coupling, $g_Z^{} = g/c_W^{}$ with $g$ being the $SU(2)_L$ gauge coupling, and $Q_\Psi$ ($T_\Psi^3$) is the electric charge (the third component of the isospin). 
The charge $X_\Psi$ is defined as 
\begin{align}
X_\Psi &= \tilde{X}_\Psi - Y_\Psi\frac{g}{g_X^{}}\tan\theta_W\tan\epsilon', \label{eq:x-charge}
\end{align}
with $\tilde{X}_\Psi$ and $Y_\Psi$ being the $U(1)_X$ charge and the  hypercharge for $\Psi$, respectively. 
The masses of the neutral gauge bosons $\tilde{Z}_\mu$ and $\tilde{Z}_\mu^\prime$ are given by the covariant derivatives of the Higgs fields $H$ and $\Phi$ by substituting their VEVs. We obtain the mass matrix in the basis of $(\tilde{Z}_\mu,\tilde{Z}_\mu^\prime)$ as 
\begin{align}
{\cal M}=\frac{1}
{4}
\begin{pmatrix}
g_Z^2 v^2 & -2g_Z^{}g_XX_H v^2 \\
-2g_Z^{}g_XX_H v^2 & 4g_X^2(X_H^2v^2 + X_\Phi^2v_\Phi^2 )
\end{pmatrix}. 
\end{align}
This can be diagonalized by introducing the orthogonal transformation as 
\begin{align}
\begin{pmatrix}
\tilde{Z}_\mu\\
\tilde{Z}'_\mu
\end{pmatrix}=
R(\zeta)
\begin{pmatrix}
Z_\mu\\
Z'_\mu
\end{pmatrix},~~
\sin 2 \zeta = \frac{g_Z^{}g_X^{}  X_H v^2}{m^2_{Z'} - m^2_Z},
\end{align}
where $m_Z^{}$ and $m_{Z^\prime}^{}$ are the mass eigenvalues of $Z$ and $Z'$, respectively, and $Z$ is identified as the discovered $Z$ boson.
The magnitude of $g_X^{}X_H$ has been severely constrained by experiments, so that we can express physical quantities by a perturbation of $g_X^{}X_H$. Under $|g_X^{}X_H| \ll 1$, we can approximate 
\begin{align}
m_Z^{} &= \frac{g_Z^{}v}{2} + {\cal O}(g_X^{2}X_H^2), \\ 
m_{Z^\prime}^{} &= g_X^{}X_\Phi v_\Phi^{} + {\cal O}(g_X^{2}X_H^2) , \label{eq:mzp}\\
\zeta &= \frac{m_Z^{}v}{m_{Z'}^2 - m_Z^2}g_X^{}X_H + {\cal O}(g_X^{3}X_H^3) . \label{eq:zeta}
\end{align}
In the dark photon limit ($\tilde{X}_\Psi \to 0$, $\Psi \neq \Phi$) with $m_{Z^\prime}^{}/m_Z^{}\ll 1$, we obtain 
\begin{align}
\zeta \to \epsilon' s_W^{} + {\cal O}(\epsilon^{\prime 3}). 
\end{align}
We can also extract the Scalar-Gauge-Gauge type interactions defined as 
\begin{align}
{\cal L}_{\rm int}& =  (Z_\mu,Z_\mu')
\begin{pmatrix}
\Lambda_{ZZ} & \Lambda_{ZZ^\prime} \\
\Lambda_{ZZ^\prime} & \Lambda_{Z^\prime Z^\prime}
\end{pmatrix}
\begin{pmatrix}
Z^\mu\\
Z^{\prime \mu}
\end{pmatrix}, 
\end{align}
where 
\begin{align}
\Lambda_{ZZ} &= \frac{m_Z^2}{v}\tilde{h} + m_{Z'}^2\left(\frac{\tilde{\phi}}{v_\Phi} - \frac{\tilde{h}}{v}\right)\zeta^2 + {\cal O}(\zeta^4), \\
\Lambda_{Z^\prime Z^\prime} &= \frac{m_{Z'}^2}{v_\Phi^{}}\tilde{\phi} - \frac{m_{Z'}^4}{m_Z^2} \left(\frac{\tilde{\phi}}{v_\Phi} - \frac{\tilde{h}}{v}\right)\zeta^2 + {\cal O}(\zeta^4), \\
\Lambda_{ZZ^\prime} &= m_{Z'}^{2}\left(\frac{\tilde{\phi}}{v_\Phi} - \frac{\tilde{h}}{v}\right)\zeta + {\cal O}(\zeta^3). \label{eq:Lambda12}
\end{align}
In the above expression, if we replace $\tilde{h} \to v$ and $\tilde{\phi} \to v_\Phi^{}$, then $\Lambda_{ZZ}$ ($\Lambda_{Z'Z'}$) becomes the mass of $Z$ ($Z'$), and $\Lambda_{ZZ'}$ vanishes as we expect.   

There are five free parameters in our scenario, which can be expressed as 
\begin{align}
\{g_X^{},~\epsilon',~m_\phi,~v_\Phi,~\sin\alpha\}.
\end{align}
Instead of $\{v_\Phi^{},\epsilon'\}$, we can choose $\{m_{Z^\prime},\zeta \}$ via Eqs.~(\ref{eq:mzp}) and (\ref{eq:zeta}) as inputs. 

\section{$Z \to Z' \phi$ decay process \label{sec:width}}

The decay width of the $Z \to Z' \phi$ mode is calculated 
from the $ZZ^\prime \phi$ coupling which is obtained by substituting $ \tilde{\phi} \to \cos\alpha$ and $\tilde{h} \to -\sin\alpha$ in Eq.~\eqref{eq:Lambda12}.
The decay width is then given by 
\begin{align}
&\Gamma(Z \to Z'\phi) = 
 \frac{m_Z^3}{48\pi }\left(\sin\alpha + \frac{v}{v_\Phi^{}}\cos\alpha \right)^2 x_{Z^\prime}^{}\zeta^2\nonumber \\
&~~~ \times \lambda^{1/2}(x_{Z'},x_\phi)\left[(1 + x_{Z'} - x_\phi)^2 + 8x_{Z'}\right],
\end{align}
where $x_{Z'} = m_{Z'}^2/m_Z^2$, $x_{\phi} = m_{\phi}^2/m_Z^2$ and $\lambda(x,y) = 1+x^2+y^2-2xy-2x-2y$.
For $x_{Z'} \ll 1$ and $x_{\phi} \ll 1$, the decay width can be simplified to be   
\begin{align}
\Gamma(Z \to Z'\phi) 
&\sim  \frac{m_Z^{3}}{48\pi v^2}\left(\sin\alpha +  \frac{v}{v_\Phi^{}}\cos\alpha\right)^2 x_{Z'}\zeta^2. \label{eq:zzpphi}
\end{align}
The branching ratio (BR) of $Z \to Z' \phi$ is given by $\text{BR}(Z \to Z' \phi) = \Gamma(Z \to Z' \phi)/\Gamma_Z$ where $\Gamma_Z$ is the total decay width of $Z$. 
When the contribution from the new decay channel (\ref{eq:zzpphi}) to $\Gamma_Z$ is negligibly small, we can use the observed value $\Gamma_Z\sim 2.50$ GeV~\cite{ParticleDataGroup:2024cfk}. 
The BR is then estimated as 
\begin{align}
\text{BR}(Z \to Z'\phi)  
& \sim 0.033\times \left(\sin\alpha + \frac{v}{v_\Phi^{}}\cos\alpha \right)^2 x_{Z'}\zeta^2. \label{eq:BR}
\end{align}
If we take the non-mixing limit $\sin\alpha \to 0$, 
the above expression becomes 
\begin{align}
\text{BR}(Z \to Z'\phi) 
& \sim 0.24 \times (g_X^{}X_\Phi{}\zeta)^2,\label{eq:BR2}
\end{align}
in which the $m_{Z^\prime}$ dependence disappears if we neglect ${\cal O}(x_{Z^\prime})$ corrections.

 \begin{figure}[t]
 \begin{center}
\includegraphics[width=9cm]{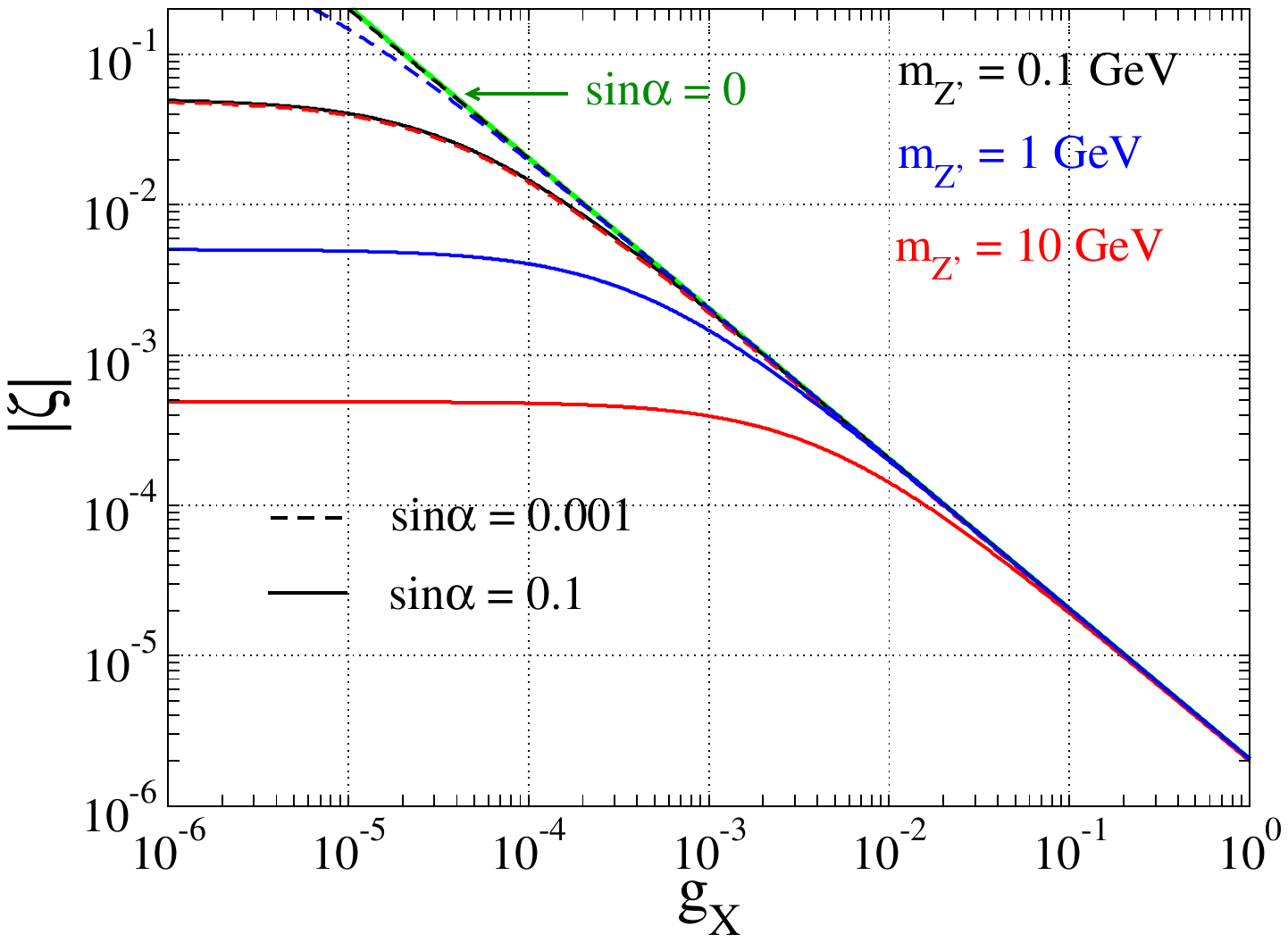}
\caption{Contour plots of BR$(Z \to Z'\phi) = 10^{-12}$ on the $\{g_X, |\zeta| \}$ plane with $\sin\alpha = 0.1$ (solid curves) and $10^{-3}$ (dashed curves). 
The value of $m_{Z^\prime}$ is taken to be 0.1 GeV (black), 1 GeV (blue) and 10 GeV (red). 
We fix $m_\phi/m_{Z^\prime} = 2.5$ and $X_\Phi = 1$. We also show the case with $\sin\alpha = 0$ as the green line using Eq.~(\ref{eq:BR2}).}
\label{fig:low-mass}
\end{center}
\end{figure}

In Fig~\ref{fig:low-mass}, we show the contours providing $\text{BR}(Z \to Z'\phi) = 10^{-12}$ on the $\{g_X, |\zeta| \}$ plane such that ${\cal O}(1)$ events are expected on each contour at Tera-Z experiments. 
We take $\sin\alpha = 0.1$ (solid curves) and $10^{-3}$ (dashed curves).
As explained above, the $m_{Z^\prime}$ dependence becomes smaller for smaller $\sin\alpha$. In fact, for $m_{Z^\prime} \leq 1$ GeV and $\sin\alpha = 10^{-3}$ the BR almost agrees with the one given by Eq.~(\ref{eq:BR2}) shown as the green line in the plot. 
On the contrary, for $\sin\alpha = 0.1$ 
the BR takes a constant value, i.e., 
$\sim 0.033\times x_{Z'}\zeta^2\sin^2\alpha $, 
when $g_X^{}$ is smaller than about $10^{-3}$, $10^{-4}$ and $10^{-5}$ with $m_{Z^\prime} = 10$ GeV, 1 GeV and 0.1 GeV, respectively. This is because the $\sin\alpha$ term inside the parentheses in Eq.~(\ref{eq:BR}) dominates compared with the second term. 

The new scalar boson $\phi$ can decay into a pair of SM particles (mostly into a fermion pair for $m_{\phi} \ll m_W^{}$) and $Z^\prime Z^\prime$. 
The ratio of the magnitudes of two decay widths is estimated as 
$\Gamma(\phi \to f\bar{f})/\Gamma(\phi \to Z'Z') \sim (y_f \tan\alpha/g_X^{})^2\times (m_\phi/m_{Z'})^2$ with $y_f$ being the SM Yukawa coupling for a fermion $f$. 
Thus,  $\phi \to Z'Z'$ is dominated for $|\sin\alpha| \ll 1$, which is typically required by various experiments~\cite{Ferber:2023iso}.
In particular, light scalar productions from meson decays constrain $|\sin\alpha| \lesssim 10^{-4}$ for $m_\phi \lesssim 5$ GeV. 
On the other hand, the constraint on $g_X^{}$ depends on the choice of the $U(1)_X$ charges. If we consider the case where 
the SM fermions are charged under $U(1)_X$, e.g., $B-L$ and $L_i - L_j$~\cite{Bauer:2018onh}, $g_X$ has severely been constrained by flavor experiments, typically $g_X^{} \lesssim 10^{-4}$ for $m_{Z'}\lesssim 10$ GeV~\cite{Bauer:2018onh}. On the contrary, if we consider the dark photon case, $g_X^{}$ is not constrained by the flavor data, but it is constrained by the Higgs boson decay into multi-dark photon states depending on the value of $\sin\alpha$~\cite{Nomura:2024pwr}.  

In the following discussion, we focus on the dark photon case ($\tilde{X}_\Psi =0$), in which we can take $g_X^{}$ to be order one for $\sin\alpha \ll 1$ and then the decay of $\phi \to Z'Z'$ becomes dominant mode.

\section{Perspective in the dark photon case \label{sec;results}}

For $\epsilon' \ll 1$, the dark photon interaction with a SM fermion $f_{}$ is written by 
\begin{equation}
\mathcal{L}_{Z'f\bar f} = e \epsilon Q_f \bar{f} \gamma^\mu f Z'_\mu,
\label{eq:DPint}
\end{equation}
where $\epsilon \simeq \epsilon' \cos \theta_W$ and $Q_f$ is the electric charge of $f$.
The parameter $\epsilon$ has been constrained by various dark photon searches~\cite{Ilten:2018crw} such as 
beam dump experiments~\cite{Riordan:1987aw,Blumlein:1991xh,NA482:2015wmo,NA64:2019auh,NA62:2023qyn}, BarBar~\cite{BaBar:2014zli,BaBar:2016sci}, LHCb~\cite{LHCb:2017trq,LHCb:2019vmc} and FASER~\cite{Petersen:2023hgm}.

 \begin{figure}[t]
 \begin{center}
\includegraphics[width=8cm]{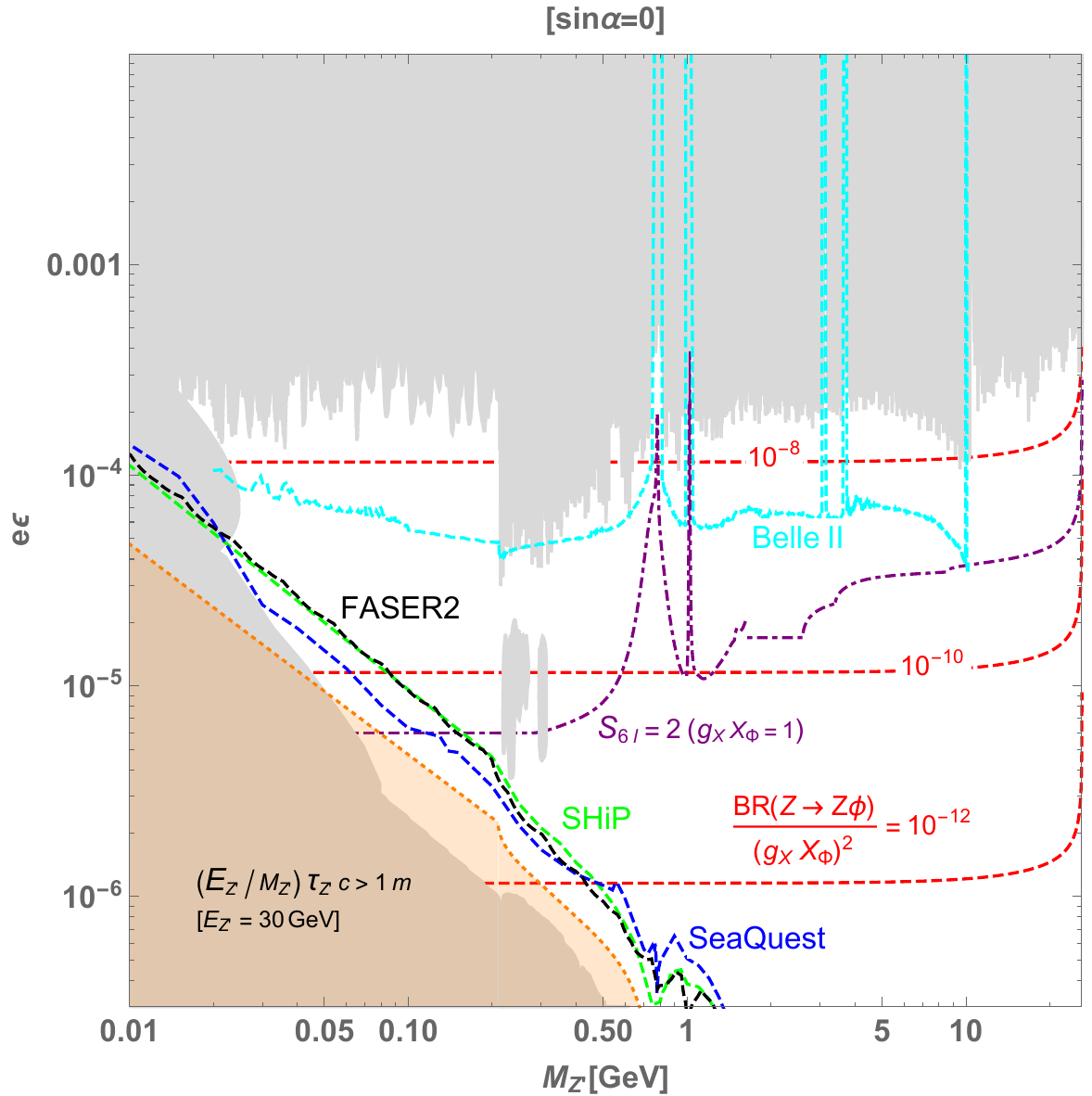}  
\caption{Red-dashed curves show contours for fixed values of the branching ratio BR$(Z \to Z'\phi)$ on the $\{m_{Z'}, e \epsilon \}$ plane with $\sin \alpha =0$ and $m_\phi = 2.5 m_{Z'}$.  The purple dot-dashed curve indicates the parameters that realize $S_{6\ell} =2$, see Eq.~(\ref{s6l}), for integrated luminosity of $10^5$ fb$^{-1}$ and $g_X X_\Phi =1$. The gray-shaded regions are excluded by current experiments. 
The bounds expected at future experiments
are also shown by the black (FASER2), green (ShiP), blue (SeaQuest) and cyan (Belle II) dashed curves. }
\label{fig:contourBR}
\end{center}
\end{figure}

In Fig.~\ref{fig:contourBR}, we show the contours giving fixed values of BR$(Z \to Z' \phi)/(g_X X_\Phi)^2$ on the $\{m_{Z'}, e \epsilon \}$ plane with $\sin \alpha =0$.
The gray shaded region is excluded by the combination of the experiments mentioned above.
We find that the significantly stronger limit on the value of $e\epsilon$ than the current limit can be obtained at Z-factories. 
In particular, the case with $e \epsilon \gtrsim 10^{-6}$ can be 
explored at Tera-Z experiments for $g_XX_\Phi = \mathcal{O}(1)$.  
In addition, we show future perspectives from other experiments for comparison: the region above the cyan-dashed curve can be tested by the Belle II experiment~\cite{Belle-II:2018jsg}, while the region below the blue, green and black dashed curves can be tested by SeaQuest~\cite{Berlin:2018pwi}, SHiP~\cite{SHiP:2015vad,SHiP:2020vbd} and FASER2~\cite{Feng:2017uoz,FASER:2018eoc} experiments respectively. Thus, our signal would have advantage, especially for the region with $m_{Z'} \gtrsim 0.03$ GeV and $ e \epsilon \gtrsim 10^{-6}$, to test the dark photon model.

The dark photon decays into SM fermions via the interaction given in Eq.~\eqref{eq:DPint}, whose BR depends on $m_{Z'}$.
For $m_{Z'} \lesssim 2 m_\mu$, the BR of $Z'\to e^+e^-$ is almost 100\%. 
For $m_{Z'} \sim 10$ GeV, the BRs of $Z'$ into a pair of charged leptons ($e^+e^-/\mu^+\mu^-/\tau^+\tau^-$) are around 15\%, while its hadronic mode is around 55\%.  
For illustration, we consider the following benchmark points (BPs);
\begin{align}
& {\rm BP1}: \ m_{Z'} = 0.1 \ {\rm GeV}, \ e \epsilon = 10^{-5}, \ g_X =0.5, \\
& {\rm BP2}: \ m_{Z'} = 10 \ {\rm GeV}, \ e \epsilon = 10^{-5}, \ g_X =0.5, 
\end{align}
where $m_\phi = 2.5 m_{Z'}$ and $X_\Phi =1$ in both BPs.
In Table~\ref{tab:1}, we show the expected number of events for some final states in two BPs under $10^{12}$ Z boson production, where ``had.'' indicates hadronic final states. In BP1, $Z'$ only decays into $e^+e^-$, so that we obtain 3 pairs of $e^+ e^-$ from the Z decay. In BP2, $Z'$ can decay into both leptons and hadronic states, and various final states are obtained. In these BPs, we would have a sufficient number of events including 
multi-lepton with or without hadronic jets, which cannot be realized by the usual decays of the Z boson in the SM. 
 In the following, we roughly discuss the SM background to estimate the signal significance at parton level.

We focus on the SM backgrounds for the ``6 lepton'' final states which mainly come from 
the $e^+e^- \to 3\times (\ell^+\ell^-)$ and $e^+e^- \to 2\times (\ell^+\ell^-)+2~\text{jets}$ processes with the latter jets being miss identified to be leptons. 
We use {\tt MadGraph5}~\cite{Alwall:2014hca} to calculate cross sections of the above processes in the SM. Imposing the basic rapidity cuts $|\eta(\ell)| < 2.5$ and $|\eta({\rm jet})|< 5$, we obtain 
\begin{align}
& \sigma(e^+ e^- \to 3\times(\ell^+\ell^-)) \simeq 2.3 \times 10^{-1} \ {\rm fb}, \\
& \sigma(e^+ e^- \to 2\times(\ell^+\ell^-)+\text{2 jets}) \simeq 3.1  \ {\rm fb}.
\end{align}
The background events can be reduced by kinematical cuts for the transverse momentum of leptons and jets since some leptons (jets) are induced from initial/final state radiations. In fact, when we apply cuts $p_T > 5$ GeV for leptons/jets the cross sections become 
\begin{align}
& \sigma(e^+ e^- \to 3\times(\ell^+\ell^-)) \nonumber \\
& \qquad \simeq 3.7 \times 10^{-4} \ {\rm fb} \ (p_T > 5 \ {\rm GeV}), \\
&  \sigma(e^+ e^- \to 2\times(\ell^+\ell^-)+\text{2 jets}) \nonumber \\
& \qquad \simeq 1.4  \times 10^{-3} \ {\rm fb} \ (p_T > 5 \ {\rm GeV}).
\end{align}
Thus, the number of background events is expected to be
\begin{equation}
N^{\rm BG}_{6 \ell} = 37, \quad N^{\rm BG}_{4\ell 2j} = 140,
\end{equation}
where we take the integrated luminosity at Tera-Z experiments to be $L = 10^{5}$ fb$^{-1}$. 
On the other hand, we find the efficiency under kinematical cuts for 6-lepton signal is $\epsilon_{\rm eff} \sim 0.14$ by generating events with {\tt MadGraph5} implementing relevant interactions.
We then estimate signal significance by 
\begin{equation}
S_{6 \ell} = \frac{N^{S}_{6\ell}}{\sqrt{N^{S}_{6\ell}+N^{\rm BG}_{6\ell} + \epsilon_{\rm mis}^2 N^{\rm BG}_{4\ell 2 j} }}, \label{s6l}
\end{equation}
where $N^S_{6 \ell}$ is number of 6-lepton signal events after kinematical cuts and $\epsilon_{\rm mis}$ is rate of jet miss-identification as charged lepton. Here we adopt $\epsilon_{\rm mis} \sim 0.01$~\cite{Yu:2021pxc}.
We then show the values of $\{M_{Z'}, e \epsilon\}$ that can realize $S^{S}_{6 \ell} =2$ for $g_X X_\Phi=1$ by purple dot-dashed curve in Fig.~\ref{fig:contourBR}.
It is thus shown that we can obtain high sensitivity from the 6-lepton signal in wide parameter region where the region above the curve could be tested in 95\% confidence level.
We can also expect to enhance significance by using other modes with optimized kinematic cuts.
Indeed, in order to estimate the feasibility of detecting the signatures more precisely, one has to perform detailed simulation studies including all the possible backgrounds such as pileup and misidentified hadrons, but these are beyond the scope of this article, which will be explored in upcoming works.

\begin{table}[t]
\begin{tabular}{c||c c c c}\hline  
&  ~$6 \ell $~ & ~$4 \ell$+had.~ & ~$2 \ell$+had.~ & ~had.~ \\\hline
\hspace{2mm}BP1\hspace{2mm}  & $46$ & $0$ & $0$ & $0$   \\
\hspace{2mm}BP2\hspace{2mm}  & $1.1$ & $6.1$ & $11$ & $6.6$ \\\hline
\end{tabular}
\caption{Expected number of events for two BPs under $10^{12}$ Z boson production, where ``had.'' indicates hadronic final states and $\ell =\{e, \mu \}$. }\label{tab:1}
\end{table}

 \begin{figure}[t]
 \begin{center}
\includegraphics[width=8cm]{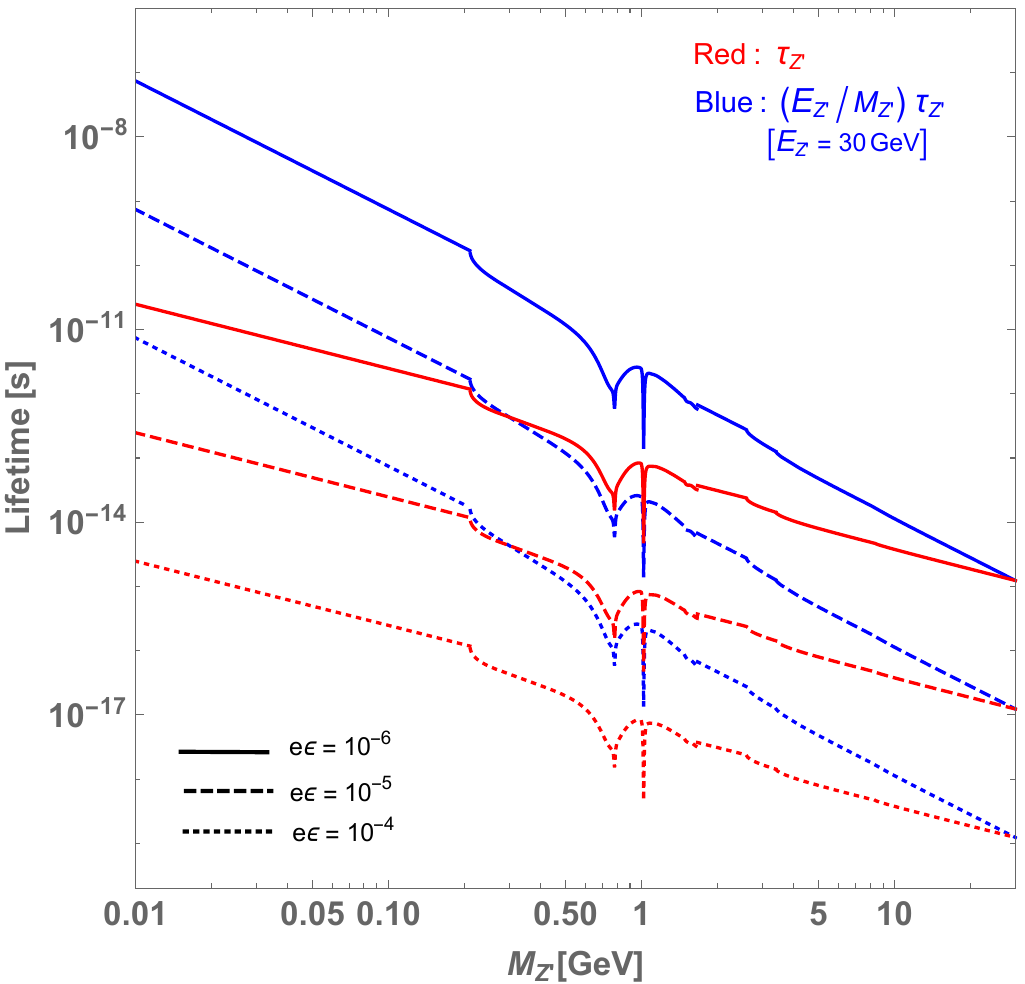}  
\caption{The lifetime of $Z'$ as a function of $Z'$ mass where solid, dashed and dotted line corresponds to $e \epsilon = 10^{-6}$, $10^{-5}$ and $10^{-4}$.  The red and blue color distinguish lifetime for rest $Z'$ and moving $Z'$ with relativistic factor of $E_{Z'}/m_{Z'}$ ($E_{Z'}=30$ GeV) }.
\label{fig:lifetime}
\end{center}
\end{figure}

Before concluding, we comment on the other possibilities of the Z boson decays than the $Z \to Z' \phi \to Z'Z'Z'$ process as follows.
\begin{itemize}
\item For $m_{Z'} < m_{\phi} < 2 m_{Z'}$, $\phi$ can decay into three body final states, i.e., $\phi \to Z'Z'^*(\to f \bar f)$ instead of an on-shell $Z'$ pair. In this case, the $\phi \to f\bar{f}$ mode can be more important compared with the case for $m_\phi = 2.5 m_{Z'}$, depending on $\sin\alpha$.  
For small $e \epsilon$ and/or $|\sin\alpha|$, 
$\phi$ becomes long-lived. 
\item For $m_\phi < m_{Z'}$, $\phi$ mainly decays into $f \bar f$ via the scalar mixing. For a small scalar mixing, $\phi$ becomes long-lived. 
\item 
$Z'$ can be long-lived when it is light as e.g., $m_{Z'} \lesssim 100$ MeV and $\epsilon$ is smaller than ${\cal O}(10^{-6})$ as indicated in Fig.~\ref{fig:lifetime}. In the collider experiment, light $Z'$ is relativistic and the lifetime is roughly longer by factor of $E_{Z'}/m_{Z'}$ with $E_{Z'}$ being $Z'$ energy.
In the figure, the red and blue color distinguish lifetime for rest $Z'$ and moving $Z'$ with relativistic factor of $E_{Z'}/m_{Z'}$ ($E_{Z'}=30$ GeV) 
Note that $Z'$ is long-lived around the gray shaded region in the left-lower area in Fig.~\ref{fig:contourBR}  indicated by orange region. 
\end{itemize}
For the long-lived $\phi$ and/or $Z'$ case, signal events include a displaced vertex that would be the good target of detector located far away from the collision point~\cite{Lu:2024fxs}.
Analyses of these cases would be more complicated than the $Z'Z'Z'$ case, and detailed simulation studies are required to estimate the discovery potential.

\section{Conclusion \label{sec:conclusion}}

We have discussed the $Z \to Z' \phi \to Z'Z'Z'$ process in models with a spontaneously broken $U(1)_X$ gauge symmetry that contains a new neutral gauge boson $Z'$ and a scalar boson $\phi$. 
Such a new decay channel of the $Z$ boson can be tested at future Z-factories, e.g., CEPC and FCC-ee, where ${\cal O}(10^{12})$ $Z$ bosons are expected to be produced. 
The branching ratio of $Z$ into the triple $Z'$ state can be larger then ${\cal O}(10^{-12})$ when 
a new gauge coupling $g_X^{}$ is larger than about $10^{-4}$ ($10^{-1}$) for the $Z$-$Z'$ mixing $\zeta\simeq 0.02$ ($2\times 10^{-5}$) and the scalar mixing $\sin\alpha = 0$. When SM fermions are charged under the $U(1)_X$ symmetry, both $g_X^{}$ and $\zeta$ have highly been constrained by the current experiments. On the other hand, $g_X^{}$ has not been severely constrained in the dark photon case, so that a sizable branching ratio of the $Z \to Z'Z'Z'$ is allowed.   
Therefore, the dark photon scenario is the most promising one which can be explored at future Z-factories.

Focusing on the dark photon case, it has been found that detectable number of events is expected for $\epsilon e \gtrsim 10^{-6}$ with $g_X = \mathcal{O}(1)$ if the decay mode is kinematically allowed. In addition, our brief estimation of signal and background shows that $e \epsilon \gtrsim \mathcal{O}(10^{-5})$ could be tested. Note also that even smaller values of $\epsilon$ and/or $g_X$ 
the process can be tested when $X_\Phi$ is larger~\cite{A:2024shl}. 
We have also shown that a sufficiently large number of events for multi-lepton with or without hadronic jets
 can be obtained in the benchmark points by considering some final states from the $Z'$ decay.
Such final states cannot be realized by the usual decay of the Z boson in the SM. 
Therefore, Z-factories provide an excellent opportunity to explore the scenarios with the $U(1)_X$ gauge symmetry. 

\acknowledgments
The work was supported by  the Fundamental Research Funds for the Central Universities (T.~N.).

\bibliography{references}

\end{document}